\documentclass{iucrjournals}

\usepackage{amsmath}
\usepackage{fancyvrb}
\usepackage{soul}

\title{Screening of tensor properties by magnetic point group symmetries: The PythMPG code package}

\author[a,b]{Andrea Urru\IUCrCemaillink{aurru@dsf.unica.it}\IUCrOrcidlink{0000-0002-5602-3063}}%
\author[c]{Turan Birol\IUCrEmaillink{tbirol@umn.edu}\IUCrOrcidlink{0000-0001-5174-3320}}%
\author[a]{Trey Cole\IUCrCemaillink{bc889@physics.rutgers.edu}\IUCrOrcidlink{0000-0002-6810-4186}}%
\author[a]{David Vanderbilt \IUCrEmaillink{dhv@physics.rutgers.edu}\IUCrOrcidlink{0000-0002-2465-9091}}

\affil[a]{Department of Physics and Astronomy, Center for Materials Theory, Rutgers University, Piscataway, NJ, 08854, USA.}
\affil[b]{Department of Physics, University of Cagliari, Cittadella Universitaria, Monserrato, CA, 09042, Italy.}
\affil[c]{Department of Chemical Engineering and
Materials Science, University of Minnesota, Minneapolis, MN, USA.}

\begin{document} 
\nolinenumbers
\maketitle 

\begin{synopsis}
We present PythMPG, a self-contained code package for symmetry-based screening of tensor properties in magnetic point groups. It provides an efficient framework for identifying symmetry-allowed tensor properties and guiding the search for materials exhibiting desired physical effects.
    
\end{synopsis}

\begin{abstract}
Understanding the symmetry requirements that permit specific phenomena or physical effects is crucial in condensed matter physics for identifying candidate materials that exhibit one or more of these effects. To this end, screening tensor properties based on (magnetic) point-group symmetries is of particular interest. Here, we present the PythMPG code package, a tool designed for this purpose. Inspired by the MTENSOR utility of the Bilbao Crystallographic Server, PythMPG performs symmetry analysis in a self-contained way to determine which tensors, characterized by their Jahn symbols, are allowed for each magnetic point group and how many symmetry-allowed independent components each tensor possesses. We describe the structure of the code and present some illustrative examples.
\end{abstract}


\section{Introduction}

Crystallographic symmetry plays a central role in the theory of materials and their physical properties.  Concepts of space group and point group -- and, for magnetic systems, magnetic space group and magnetic point group (MPG) -- are foundational in this context. A knowledge of the (magnetic) space group is required for many purposes, such as scattering properties and dispersion relations. However, the presence or absence of macroscopic properties such as electric polarization, linear and nonlinear dielectric response, magnetization, and optical properties are controlled by the symmetry of the point group.

Recent years have seen an acceleration of a trend toward the study of novel magnetic materials, especially those for which the presence or absence of some symmetry plays a crucial part. The roles of broken inversion and time-reversal symmetries, chirality, and other details of the MPG for activating novel phenomena are under intense investigation for the development of improved materials for a variety of applications. In particular, altermagnetism, spintronics, magnetic topological materials, chirality induced spin selectivity, and nonlinear optics are all very active fields in which symmetry principles play a crucial role \cite{Mazin2022, Jungwirth2026, Baltz2018, Zutic2004, Tokura2019, Bloom2024, Fiebig2004}. Purported loop-current phases, in which the magnetism does not arise in the spin sector, can also be classified using magnetic groups \cite{fernandes_loop-current_2026}.

Here we describe a Python code package that provides tools for analyzing the allowed physical properties of materials based on the symmetries encoded in the MPG.  While the emphasis is on magnetic materials for the reasons sketched above, the 32 ordinary crystallographic point groups are just special cases of the 122 MPGs, and thus are covered as well.
Inspired by the MTENSOR utility of the Bilbao Crystallographic Server~\cite{gallego-actacrysta19}, PythMPG performs symmetry analysis in a similar but entirely independent and self-contained fashion. It is designed to allow a user to determine which tensors, characterized by their Jahn symbols, are allowed for each magnetic point group, and if allowed, how many independent tensor components are present.


\subsection{Symmetry groups}

The 32 crystallographic point groups (PGs) enumerate the possible symmetry groups of proper and improper rotations that bring a three-dimensional crystal back onto itself, up to translations, and modulo equivalence of the entire symmetry group under global rotation \cite{Bradley2023}. Proper rotations are those (including the identity) that retain the right-handed sense of the Cartesian frame. Improper rotations are those, like inversion itself, mirror operations, and rotoreflections, that reverse the sense of the Cartesian frame. The PGs are the natural symmetry groups for describing nonmagnetic systems.

For magnetic crystals, the appropriate generalization of the PG is the magnetic PG (MPG).  In this case the symmetry operations include proper and improper rotations as before, but they may also include those same operations composed with time reversal (TR) \cite{Borovik2006}. The 32 PGs reappear as 32 MPGs, known as colorless groups, in which TR does not appear in any operation. They also reappear in doubled form as 32 nonmagnetic MPGs, known as gray groups, in which TR itself is a symmetry, so that every proper or improper rotation occurs both with and without TR. Finally there are 58 MPGs, known as black-white groups, in which half of the symmetry operations occur without TR and the other half with TR, but TR itself is not a symmetry. We focus here on the 122 MPGs, with the understanding that PGs can be recovered as special cases of MPGs.

The MPGs can also be used to characterize molecules and clusters, except in that case the number of allowed PGs and MPGs is enlarged by the possibility of symmetry operations, such as five-fold rotations, that are incompatible with the translational symmetry of crystalline lattice structures. 

When translations, both primitive and non primitive, are combined with the PG symmetry operations, one arrives at the 230 space groups, each of which is uniquely associated with a PG. Similar considerations lead to 1651 possible magnetic space groups.  The space groups will not concern us further here.

A reason for focusing on MPGs is that they determine the possible macroscopic order parameters and response functions allowed by symmetry \cite{gallego-actacrysta19, Dayroberts2025}.  This is a consequence of Von Neumann's principle, which states that the physical macroscopic properties of a crystal have to respect the symmetries of the MPG. To take a simple example, a four-fold rotational symmetry about an axis constrains both the electric polarization and spin magnetization to lie on the rotational axis. Instead, a time-reversed four-fold rotation still constrains the electric polarization to lie along the axis, but forces the magnetization to vanish completely.  Higher-order tensor properties such as linear and nonlinear dielectric response, piezoelectricity, piezomagnetism, and anomalous Hall conductivity, are similarly constrained by the MPG symmetries. 

We shall use the term ``response tensor'' to refer to such physical properties, also including vector quantities like electric polarization (viewed as rank-1 tensors) and even a few scalar properties (viewed as rank-0 tensors). Clearly, it is useful to be able to test whether a particular tensor quantity is symmetry-allowed for a crystal belonging to a particular PG or MPG.  This is precisely the purpose of the TENSOR and MTENSOR modules, respectively, of the Bilbao Crystallographic Server \cite{Aroyo-ACA2006, Aroyo-zkri2006, Aroyo-bcc2011, gallego-actacrysta19}. In this browser-based online resource, one can pose exactly this kind of query.  In case the answer is affirmative, the module will also specify the number of independent tensor components, and indicate the symmetry-allowed forms in which these components are arranged in some standard crystallographic setting.

\subsection{Design of the PythMPG package}

The present project aims to provide precisely the same kind of information, and indeed in many respects is modeled after the MTENSOR tool.  For example, we adopt the same convention of describing response tensors by their ``Jahn symbols'' as described in the next section.  However, it differs in the following respects.
\begin{itemize}
\item The PythMPG package is an independent, stand-alone Python package released under the GPL-3.0 public license. No crystallographic database information was used in the code development, other than the conventional ordered list of MPG names. Once installed, the code executes with no web access required.
In this respect, it is an example of the increasing number of Python-based crystallographic symmetry tools such as Spglib for Python \cite{Togo2024}, IrRep \cite{Iraola2022}, ProDenCer \cite{prodencerGitHub}, or pymatgen.symmetry \cite{Ong2013}. 
\item By default the code operates by taking a list of Jahn symbols as input, and constructing a comma-separated-value (CSV) file in which all 122 MGPs appear as rows, and a chosen list of tensor Jahn symbols appear as columns.  This delivers a large volume of information simultaneously, and allows for selection and sorting using spreadsheet methods.
\item The present code is capable of reporting the number of independent components of each tensor under each MPG, but does not describe the shape of the allowed tensor in the same way that MTENSOR does.
\item The Python code takes the form of five modules that are designed to work together, but that could be reused independently for other purposes.  For example, one module performs the parsing of Jahn symbols, and another constructs libraries of proper crystallographic rotations and their multiplication tables.
\item The design and coding is completely independent of that of
MTENSOR, allowing for cross-checking between the codes.
\end{itemize}

\subsection{MPG terminology}

Several systems are in common use for naming the MPGs, of which the Hermann-Mauguin (HM) short names appear to be most commonly used.  There is also a system of HM long names which sometimes differ from the short names, as for example $2/m\,2/m'\,2/m'$ (long) as a synonym for $mm'm'$ (short). There are alternative Schoenflies and Shubnikov notations, in which $mm'm'$ appears as $D_{2h}(C_{2h})$ and $\underline{m}\!\cdot\!2\!:\!m$ respectively. Neither these nor the HM long-form names are currently supported by the PythMPG package.

Even within the HM short-name system, variations occur.  A good summary appears in Table S1 in the Supplemental Material of \cite{campbell-acsa22}, where ``old'' MPG short names are listed along-side proposed ``UNI'' names.\footnote{That list also includes ``Litvin'' sequence numbers, which appear in the same sequence in PythMPG.}  These are identical for black-white groups, but the latter include a final ``.1'' suffix for colorless groups (compare $2/m.1$ vs.\ $2/m$) and an extra period for grey groups (compare $2/m.1'$ vs.\ $2/m1'$). With the exception that we use $1'$ to denote the group they label as $11'$, we adopt the ``old'' names from that table. This choice constitutes what are called ``standard'' MPG names below and throughout the PythMPG package.

Other choices are possible.  This usually corresponds to a different choice of setting, leading to the distinction, for example, between $mm'm'$ and our $m'm'm$. Aside from a few such variations in setting, our standard nomenclature is the same as that in the Magnetic Point Groups section of the Wikipedia \href{https://en.wikipedia.org/wiki/Magnetic_space_group}{Magnetic Space Groups} page.\footnote{Earlier versions of that page incorrectly dropped the overbars on the entries for groups $\bar{4}3m$, $\bar{4}3m1'$ and $\bar{4}'3m'$.}
For some cubic groups, there is a more subtle equivalence, such as between groups $m3m$ and $m\bar{3}m$ (the first $m$ implies the presence of three Cartesian mirrors whose product is inversion).

The PythMPG code can accept commonly used alternative names of the kinds mentioned above. However, if provided by a user in a function call, these are translated to our standard names, a warning is issued, and the results are reported using the translated names.

\subsection{Jahn symbols}


The rank of a tensor, together with its symmetry properties, are identified with a compact notation, known as the \textit{Jahn symbol} \cite{jahn-actacryst49}. The elements that make up a Jahn symbol are the following:
\begin{itemize}
    \item A number of ``$V$'' characters equal to the rank of the tensor. For tensors of rank higher than 1, the ``$V$'' characters are written in a compact way using a power-like notation. For example, $V^2$ is a rank-2 tensor, $V^3$ is a rank-3 tensor, and so on. More complicated cases exist when additional characters appear, such as for the rank-3 tensor $a[V^2]V$, as explained below.

    \item Square brackets ``[\,]'' and curly brackets ``\{\}'' identify symmetrization and antisymmetrization under permutation of the enclosed group of indices, respectively. (Note that this convention is reversed with respect to quantum commutators and anticommutators.) As an example, $[V^2]$ and $\{V^2\}$ are rank-2 tensors symmetric ($T_{ij} = T_{ji}$) and antisymmetric ($T_{ij} = -T_{ji}$) under exchange of their two indices, respectively. Similarly, Jahn symbol $[V^2]V$ identifies a rank-3 tensor symmetric under exchange of its first two indices, i.e., $T_{ijk} = T_{jik}$. Symmetrization and antisymmetrization can also apply to larger sets of indices. For example, $[V^4]$ identifies a rank-4 tensor that is symmetric under exchange of any pair of its four indices. 
    \item Characters ``$a$'' and ``$e$'' describe the transformation properties of a tensor under time-reversed and improper rotations respectively. Specifically: 
    \begin{itemize}
        \item ``$a$'' identifies a tensor that is odd under time reversal. If no ``$a$'' is present, the tensor is even under time reversal. 
        \item ``$e$'' identifies an \textit{axial} tensor. If no ``$e$'' is present, the tensor is \textit{proper}. The terms proper and axial refer to how the tensor transforms under improper point symmetry operations $\mathcal{S}$, i.e., those with $\det \mathcal{S}=-1$. In general, a proper rank-$n$ tensor transforms under $\mathcal{S}$ as
        \begin{equation}
            T_{ijk \dots z} = \mathcal{S}_{ii'} \mathcal{S}_{jj'} \mathcal{S}_{kk'} \dots \mathcal{S}_{zz'} T_{i'j'k' \dots z'},
        \end{equation}
        (implied sum over indices) whereas an axial tensor transforms as
        \begin{equation}
            T_{ijk \dots z} = (\det \mathcal{S}) \mathcal{S}_{ii'} \mathcal{S}_{jj'} \mathcal{S}_{kk'} \dots \mathcal{S}_{zz'} T_{i'j'k' \dots z'}.
        \end{equation}
        The two expressions are the same for proper symmetry operations ($\det \mathcal{S} = 1$), but differ in sign for improper symmetry operations ($\det \mathcal{S} = -1$). 
    \end{itemize}
    As an example, $V$ is a time-even proper vector (e.g., position $\mathbf{r}$), $aV$ is a time-odd proper vector (e.g., velocity $\mathbf{v}$), $eV$ is a time-even axial vector (e.g., torque $\boldsymbol{\tau} = \mathbf{r} \times \mathbf{F}$), and $aeV$ is a time-odd axial vector (e.g., magnetic field $\mathbf{B}$).
\end{itemize}
    
Jahn symbols are given for many tensor properties of interest in the Tables in \cite{gallego-actacrysta19}. Others can be found by navigating to the \texttt{MTENSOR} utility in the ``Magnetic Symmetry and Applications'' section of the \href{https://cryst.ehu.es}{Bilbao Crystallographic Server} and clicking on the links for ``equilibrium,'' ``optical,'' ``nonlinear optical susceptibility,'' and ``transport'' tensors.
At least for equilibrium tensors, it is usually also straightforward to construct the Jahn symbol by hand based on an understanding of how the tensor should transform under exchange of indices as well as under inversion and time reversal.

When dissipative processes come into play, however, extra care is needed to understand the index exchange rules in view of the Onsager reciprocal relations \cite{onsager_reciprocal_1931}. 
In \cite{gallego-actacrysta19}, the authors introduced a generalized form of Jahn symbol in which an asterisk can appear to reflect the consequences of the Onsager relations as they apply to some nonequilibrium and optical responses. For example, electrical conductivity is represented by Jahn symbol $[V^2]^*$, where the `*' indicates an interchange of a pair of indices only under time-reversed MPG operations. The current implementation of PythMPG does not recognize these extended Jahn symbols. However, the corresponding information can be obtained by an independent treatment of the symmetric and antisymmetric parts of a response tensor, as advocated by \cite{grimmer_thermoelectric_2017}. For the electrical conductivity $[V^2]^*$, for example, the symmetric (dissipative) and antisymmetric (non-dissipative) parts are described by Jahn symbols  $[V^2]$ and $a\{V^2\}$ respectively. A fairly comprehensive list of decompositions of this kind appears in Tables 1 and 2 of \cite{gallego-actacrysta19}.

\section{Availability and capabilities of the code}

\subsection{Installation}

The PythMPG package is available at PyPI, so that it can be installed just by issuing
\begin{quote}
\begin{Verbatim}[fontfamily=helvetica]
pip install pythmpg
\end{Verbatim}
\end{quote}
The full code package is available from its GitHub repository \href{https://github.com/pythmpg/pythmpg}{github.com/pythmpg/pythmpg}, which is also available in the Zenodo repository \href{https://zenodo.org/records/18672613}{zenodo.org/records/18672613} in the form of a \texttt{.zip} archive. More detailed installation instructions are provided in the repository. Documentation of the code package is hosted at \href{https://pythmpg.readthedocs.io/en/latest/}{pythmpg.readthedocs.io}.

\subsection{Direct access to tensor symmetry information}
\label{sec:getdirect}

Direct access to the symmetry information for a specified list of MPGs is provided by two functions from the \texttt{mpg\_tools} module.

Function \texttt{get\_mpg\_info()} takes a list of MPG names as input, and returns a dictionary with the following keys, for each of which the corresponding values are lists corresponding to the input list of MPG names:
\begin{itemize}
    \item \texttt{BNS}: BNS (or Litvin) serial number of each MPG.
    \item \texttt{order}: Order of the group (number of symmetry elements)
    \item \texttt{group\_type}: Colorless, gray, or black-white.
    \item \texttt{symm\_info}: Whether the group contains three basic symmetry operations (inversion $\bar{1}$, time reversal $1'$, and their combination $\bar{1}'$), each by itself or in combination with a proper rotation. 
\end{itemize}
Here is a simple script making use of this capability:
\begin{quote}
\begin{verbatim}
from pythmpg import get_mpg_info
info = get_mpg_info(["4/mmm", "m-3m"])
print(info["order"])
\end{verbatim}
\end{quote}
In this case, the result is ``[16, 48]''.

Function \texttt{get\_num\_indep()} takes as input two lists, a list of Jahn symbols and a list of MPG names.  For each pair taken from these lists, it counts the number of independent tensor components. The output is in the form of a dictionary whose keys are the Jahn symbols, and values are lists of integers corresponding to the list of MPG names. Again, here is a simple example:
\begin{quote}
\begin{verbatim}
from pythmpg import get_num_indep
count_dict = get_num_indep(["[V2]","{V2}"], ["1", "m", "4/mmm"])
for key in count_dict:
    print (f"{key} : {count_dict[key]}")
\end{verbatim}
\end{quote}
The result is:
\begin{quote}
\begin{verbatim}
[V2] : [6, 4, 2]
{V2} : [3, 1, 0]
\end{verbatim}
\end{quote}

In both of these functions, the MPG list argument is optional; if omitted, all 122 MPGs are treated.

\subsection{Extracting symmetry information in spreadsheet form}
\label{sec:getspread}

The \texttt{spreadsheet} module of the PythMPG package provides the capability of outputting a spreadsheet, in the form of a CSV (comma-separated value) file, in which the 122 MPGs correspond to rows. Some of the columns indicate the presence or absence of certain symmetries, while the remainder correspond to a user-supplied list of tensor responses as labeled by their Jahn symbols.  For those, the cells in the spreadsheet indicate the number of independent tensor components for each MPG/Jahn-symbol pair.\footnote{There is also an option to print 0 if the tensor vanishes by symmetry and 1 otherwise.} Once again, here is a simple example:
\begin{quote}
\begin{verbatim}
from pythmpg import Spreadsheet
# Initialize default spreadsheet and add two tensor operators
my_sheet = Spreadsheet()
tensors = [ ("Piezoel","V[V2]"), ("Piezomag","ae[V2]V") ]
my_sheet.add("tensor", tensors)
# Print a summary of the header sections
my_sheet.header_report()
# Build the spreadsheet and write it to disk
my_sheet.build_csv()
my_sheet.write_csv("my_mpg.csv",kind="num")
\end{verbatim}
\end{quote}

Further explanation and guidance on the usage of the \texttt{Spreadsheet} class for the construction of spreadsheets can be found in the PythMPG package documentation and in the User Guide. The latter also explains how the CSV file can be used as a basis for creating a more aesthetically formatted Excel, LibreOffice, or Numbers spreadsheet.  Once in spreadsheet form, features such as sorting and hiding of columns and rows can be used efficiently to search for MPGs that satisfy user-specified criteria.

\subsection{Access to underlying modules}

The PythMPG package takes the form of five modules that are designed to work together, but that could be reused independently for other purposes.  For example,  the \texttt{pg\_elements} module contains functions for constructing dictionaries of proper rotations and their multiplication tables; the \texttt{mpg\_dicts} module parses MPG names and associates a set of symmetry information to each; and the \texttt{parse\_jahn} module parses Jahn symbols. These capabilities are described further in the next section.

\section{Description of the code}

In brief, the code starts with a predifined list of 122 MPGs (or a user-supplied list) and parses them to create a list of polycyclic generators for each group. Some preliminary information is then extracted, such as the order or the group, its type (colorless, gray, or black-wite), and the presence or absence of $1'$, $\bar{1}$, or $\bar{1}'$ symmetries, alone or together with proper rotations. Each Jahn symbol is then parsed into a recursive data structure containing its corresponding symmetrization instructions. The (anti-)symmetrization of indices implied by `[\,]' and `\{\}' instructions is performed in turn on a complete set of unit tensors, and the resulting independent tensors are collected. Then, for each MPG in turn, these tensors are further symmetrized under the operations of the MPG to determine the number of independent tensors for that MPG. When invoked from the \texttt{Spreadsheet} class, the collected information is exported in the form of a CSV file that can be cut and pasted into a spreadsheet application such as Excel, optionally reporting the number of independent tensor components (numerical output) or simply whether this number is nonzero (Boolean output).

The package is composed of five modules as follows:

\begin{itemize}
\item {\texttt{spreadsheet}}\\
Interfaces with user-level scripts for generating MPG spreadsheets. It defines the \texttt{Spreadsheet} class, which is designed to be invoked at the user level as in the example in Sec.~\ref{sec:getspread}. This is usually the only module that needs to be accessed by the user when constructing spreadsheets. This module calls lower-level functions provided by other modules in the package as described below.

\item {\texttt{mpg\_tools}}\\
Provides two externally callable functions for querying magnetic point group properties for all 122 MPGs, or optionally, for a specified list of MPGs. These functions are also called internally from \texttt{Spreadsheet} class methods. Function \texttt{get\_mpg\_info()} returns basic symmetry information for each MPG, such as whether it contains $\bar{1}$, $1'$, or $\bar{1}'$, alone or in combination with a simple proper rotation. 

Function \texttt{get\_num\_indep()}, which is a centerpiece of the implementation, also takes a list of Jahn symbols; this time it returns the number of independent components for the corresponding tensor property in each MPG, and for each Jahn symbol. For a rank-$R$ tensor, it does this by constructing a list of $3^R$ ``unit tensors'' (1 in one location and zero elsewhere) and symmetrizing each in turn using the polycyclic generators from module \texttt{mpg\_dicts} (see below). As it proceeds, it discards any resulting zero tensors and orthogonalizes against previously collected tensors. At the end of this process, the number of remaining independent tensors (i.e., the number of independent tensor components) is reported.\footnote{For
   more detailed information, such as the tensor form showing symmetry-related tensor elements, the user is referred to
   the \texttt{MTENSOR} tool of the Bilbao Crystallographic Server.}
This module has dependencies on lower-level modules \texttt{parse\_jahn} and \texttt{mpg\_dicts}, and in turn, on \texttt{pg\_elements}. 

\item {\texttt{parse\_jahn}}\\
Defines two functions intended to be called externally to handle the parsing of Jahn symbols, or more precisely, their spatial part.  That is, Jahn symbols passed to these functions should lack any preceding 'a' or 'e' characters.\footnote{The
  application of a possible sign reversal when applying a polycyclic generator to a tensor, which depends on these characters, is handled in higher-level module \texttt{mpg\_tools}.}
Function \texttt{parse\_jahn\_symbol()} parses Jahn symbols into structured instruction blocks describing the needed symmetrization or antisymmetrization over tensor axes, and \texttt{jahn\_rank()} returns the rank (number of axes) of the tensor. Results are cached in the module-level \texttt{jahn\_dict} to avoid redundant parsing of repeated symbols.

\item {\texttt{mpg\_dicts}}\\
Constructs dictionaries that are used by higher-level modules. A pair of these define the proper rotations that appear in hexagonal and cubic frames, and another pair stores their multiplication tables. These are constructed by calling functions in module \texttt{pg\_elements} and are then made available as attributes of module \texttt{mpg\_dicts}. Of the remaining three dictionaries, one maps alternative MPG names to standard MPG names, and another maps the MPG names to BNS (Litvin) sequence numbers.

Finally, dictionary \texttt{mpg\_dict} is constructed by a call to internal function \texttt{get\_mpg\_dict()}, which defines the list of standard MPG names and then calls internal function \texttt{parse\_mpg()} on each.  The latter is a second centerpiece of the implementation, as it parses the specified MPG name and assigns to it a set of polycyclic generators~\cite{cannon-jsc04}. This means that each element of the MPG appears once and only once in a list of the form
$g_1^{k_1}\,g_2^{k_2} \,...\,g_M^{k_M}$,
where $g_m$ is the $m$'th polycyclic generator, $N_m$ is the order of that generator, and $k_m$ ranges from 0 to $N_m - 1$. Such an assignment is not straightforward, as the number $M$ of assigned generators can sometime be smaller or larger than the number of symmetry operators parsed from the MPG name. In fact, the ability to assign polycyclic generators to all crystallographic MPGs is a nontrivial result, as it is not true of all finite groups,\footnote{As explained
  in \cite{cannon-jsc04}, a finite group can be assigned a polycyclic generating sequence if and only if it is solvable, i.e., can be broken down into a sequence of nested subgroups where every quotient group is abelian.}
failing, for example, for the point group of the regular icosahedron. Polycyclic generators are used in preference to the ``generators'' listed in the International Tables or the Bilbao Crystallographic server because they make subsequent operations more straightforward and efficient. In particular, tensor symmetrization can be performed under each generator in turn with a total of $\sum_m N_m$ operations, without having to apply all $\prod_m N_m$ elements of the symmetry group.

\item {\texttt{pg\_elements}}\\
Provides tools for constructing lists of proper rotations belonging to cubic and hexagonal symmetry groups.
Function \texttt{get\_cub\_table()} constructs the elements of point group 432 and its multiplication table, while \texttt{get\_hex\_table()} does the same for group 622. These two parent groups are sufficient in that each MPG can be assigned to one of them, meaning that all of its elements can be written as elements of the parent group, either alone or in combination with inversion and/or time reversal. These functions are called from the higher-level \texttt{mpg\_dicts} module.

\end{itemize}

\section{Summary}

The PythMPG package provides much of the same functionality as the \texttt{MTENSOR} utility of the Bilbao Crystallographic Server, but takes the form of an independently implemented Python package that is designed to scan through all MPGs at once. The user provides a list of Jahn symbols describing tensor properties of interest, and the code reports the number of independent tensor components allowed by symmetry under each MPG. This information is easily made available in the form of a spreadsheet, which can be searched, sorted, or rearranged offline. The parsing of Jahn symbols to arbitrary depth, and the use of polycyclic generators to represent MPGs, are notable features of the Python implementation. It is hoped that the package will find wide utility by facilitating the guided search for materials exhibiting desired equilibrium, transport, optical, and other physical response properties.


\begin{funding}
DV and TC acknowledge support from NSF Grant DMR-2421895. AU acknowledges support from the Abrahams Postdoctoral Fellowship of the Center for Materials Theory at Rutgers University and Office of Naval Research Grant No.~N00014-21-1-2107. TB is supported by the NSF CAREER grant DMR-2046020.
\end{funding}

\ConflictsOfInterest{There are no conflicts of interest.}

\DataAvailability{The PythMPG source code is hosted on GitHub at \href{https://github.com/pythmpg/pythmpg}{\texttt{github.com/pythmpg/pythmpg}}, where the latest release and pre-release development versions can be obtained. 
Full documentation, including installation instructions, a user guide, and the application programming interface (API) reference, is available on Read the Docs at \href{https://pythmpg.readthedocs.io/en/latest/}{\texttt{pythmpg.readthedocs.io}}.
Archived snapshots of tagged releases are also deposited at
Zenodo (DOI:~\href{https://doi.org/10.5281/zenodo.18672613}{10.5281/zenodo.18672613}).}

\bibliography{iucr} 

@article{gallego-actacrysta19,
author = "Gallego, S. V. and Etxebarria, J. and Elcoro, L. and Tasci, E. S. and Perez-Mato, J. M.",
title = "{Automatic calculation of symmetry-adapted tensors in magnetic and non-magnetic materials: a new tool of the Bilbao Crystallographic Server}",
journal = "Acta Crystallographica Section A",
year = "2019",
volume = "75",
number = "3",
pages = "438--447",
month = "May",
doi = {10.1107/S2053273319001748},
url = {https://doi.org/10.1107/S2053273319001748},
}

@article{jahn-actacryst49,
       author = {{Jahn}, H. A.},
        title = "{Note on the Bhagavantam–Suranarayana method of enumerating the physical constants of crystals}",
      journal = {Acta Crystallographica},
         year = 1949,
        month = "March",
       volume = {2},
        pages = {30-33},
          doi = {10.1107/S0365110X49000060},
       adsurl = {https://doi.org/10.1107/S0365110X49000060},
}

@ARTICLE{Aroyo-bcc2011,
	author = {Aroyo, M. I. and Perez-Mato, J. M. and Orobengoa, D. and Tasci, E. and De La Flor, G. and Kirov, A.},
	title = {Crystallography online: Bilbao crystallographic server},
	year = {2011},
	journal = {Bulg. Chem. Commun.},
	volume = {43},
	number = {2},
	pages = {183 – 197},
	url = {https://www.scopus.com/inward/record.uri?eid=2-s2.0-80955140447\&partnerID=40\&md5=488772b9e21d2636a3952f66ae80ae84},
	type = {Article},
	publication_stage = {Final},
	source = {Scopus}
}

@article{Aroyo-zkri2006,
url = {https://doi.org/10.1524/zkri.2006.221.1.15},
title = {Bilbao Crystallographic Server: I. Databases and crystallographic computing programs},
author = {Aroyo, M. I. and Perez-Mato, J. M. and Capillas, C. and Kroumova, E. and Ivantchev, S. and Madariaga, G. and Kirov, A. and Wondratschek, H.},
pages = {15--27},
volume = {221},
number = {1},
journal = {Z. Kristallogr.},
doi = {doi:10.1524/zkri.2006.221.1.15},
year = {2006},
lastchecked = {2024-06-26}
}

@article{Aroyo-ACA2006,
author = "Aroyo, M. I. and Kirov, A. and Capillas, C. and Perez-Mato, J. M. and Wondratschek, H.",
title = "{Bilbao Crystallographic Server. II. Representations of crystallographic point groups and space groups}",
journal = "Acta Crystallogr. Sect. A",
year = "2006",
volume = "62",
number = "2",
pages = "115--128",
month = "Mar",
doi = {10.1107/S0108767305040286},
url = {https://doi.org/10.1107/S0108767305040286},
}

@article{campbell-acsa22,
author = "Campbell, Branton J. and Stokes, Harold T. and Perez-Mato, J. Manuel and Rodr{\'\i}guez-Carvajal, Juan",
title = "{Introducing a unified magnetic space-group symbol}",
journal = "Acta Crystallographica Section A",
year = "2022",
volume = "78",
number = "2",
pages = "99--106",
month = "Mar",
doi = {10.1107/S2053273321012912},
url = {https://doi.org/10.1107/S2053273321012912},
}

@article{cannon-jsc04,
  title={Special polycyclic generating sequences for finite soluble groups},
  author={Cannon, John J and Eick, Bettina and Leedham-Green, Charles R},
  journal={Journal of Symbolic Computation},
  volume={38},
  number={5},
  pages={1445--1460},
  year={2004},
  publisher={Elsevier}
}

@article{Baltz2018,
  title = {Antiferromagnetic spintronics},
  author = {Baltz, V. and Manchon, A. and Tsoi, M. and Moriyama, T. and Ono, T. and Tserkovnyak, Y.},
  journal = {Rev. Mod. Phys.},
  volume = {90},
  issue = {1},
  pages = {015005},
  numpages = {57},
  year = {2018},
  month = {Feb},
  publisher = {American Physical Society},
  doi = {10.1103/RevModPhys.90.015005},
  url = {https://link.aps.org/doi/10.1103/RevModPhys.90.015005}
}

@article{Zutic2004,
  title = {Spintronics: Fundamentals and applications},
  author = {\ifmmode \check{Z}\else \v{Z}\fi{}uti\ifmmode \acute{c}\else \'{c}\fi{}, Igor and Fabian, Jaroslav and Das Sarma, S.},
  journal = {Rev. Mod. Phys.},
  volume = {76},
  issue = {2},
  pages = {323--410},
  numpages = {0},
  year = {2004},
  month = {Apr},
  publisher = {American Physical Society},
  doi = {10.1103/RevModPhys.76.323},
  url = {https://link.aps.org/doi/10.1103/RevModPhys.76.323}
}

@article{Tokura2019,
  title={Magnetic topological insulators},
  author={Tokura, Yoshinori and Yasuda, Kenji and Tsukazaki, Atsushi},
  journal={Nature Reviews Physics},
  volume={1},
  number={2},
  pages={126--143},
  year={2019},
  publisher={Nature Publishing Group UK London}
}

@article{Mazin2022,
	title = {Editorial: {Altermagnetism}—{A} {New} {Punch} {Line} of {Fundamental} {Magnetism}},
	volume = {12},
	issn = {2160-3308},
	shorttitle = {Editorial},
	url = {https://link.aps.org/doi/10.1103/PhysRevX.12.040002},
	doi = {10.1103/PhysRevX.12.040002},
	number = {4},
	urldate = {2025-09-25},
	journal = {Physical Review X},
	author = {Mazin, Igor and {The PRX Editors}},
	month = dec,
	year = {2022},
	pages = {040002},
}

@article{Jungwirth2026,
	title = {Symmetry, microscopy and spectroscopy signatures of altermagnetism},
	volume = {649},
	issn = {0028-0836, 1476-4687},
	url = {https://www.nature.com/articles/s41586-025-09883-2},
	doi = {10.1038/s41586-025-09883-2},
	number = {8098},
	urldate = {2026-09-22},
	journal = {Nature},
	author = {Jungwirth, Tomas and Sinova, Jairo and Fernandes, Rafael M. and Liu, Qihang and Watanabe, Hikaru and Murakami, Shuichi and Nakatsuji, Satoru and Šmejkal, Libor},
	month = jan,
	year = {2026},
	pages = {837--847},
}

@article{Bloom2024,
    author = {Bloom, Brian P. and Paltiel, Yossi and Naaman, Ron and Waldeck, David H.},
    title = {Chiral Induced
Spin Selectivity},
    journal = {Chemical Reviews},
    volume = {124},
    number = {4},
    pages = {1950-1991},
    year = {2024},
    month = {02},
    issn = {0009-2665},
    doi = {10.1021/acs.chemrev.3c00661},
    url = {https://doi.org/10.1021/acs.chemrev.3c00661},
    eprint = {https://pubs.acs.org/chreay/article-pdf/124/4/1950/1514346/cr3c00661.pdf},
}

@article{Fiebig2004,
title = {Nonlinear optics—a powerful tool for the investigation of magnetic structures},
journal = {Journal of Magnetism and Magnetic Materials},
volume = {272-276},
pages = {E1607-E1611},
year = {2004},
note = {Proceedings of the International Conference on Magnetism (ICM 2003)},
issn = {0304-8853},
doi = {https://doi.org/10.1016/j.jmmm.2003.12.666},
url = {https://www.sciencedirect.com/science/article/pii/S0304885303023989},
author = {M. Fiebig and R.V. Pisarev},
}

@book{Bradley2023,
  title={The Mathematical Theory of Symmetry in Solids: Representation Theory for Point Groups and Space Groups},
  author={Bradley, C.J. and Cracknell, A.P.},
  isbn={9780191702334},
  series={Oxford classic texts in the physical sciences},
  url={https://books.google.com/books?id=OXRT0AEACAAJ},
  year={2023},
  publisher={Oxford University Press}
}

@incollection{Borovik2006,
  title={Magnetic properties},
  author={Borovik-Romanov, AS and Grimmer, H},
  booktitle={International Tables for Crystallography Volume D: Physical properties of crystals},
  pages={105--149},
  year={2006},
  publisher={Springer}
}

@article{Dayroberts2025,
  title = {Piezoresistivity as a Fingerprint of Ferroaxial Transitions},
  author = {Day-Roberts, Ezra and Fernandes, Rafael M. and Birol, Turan},
  journal = {Phys. Rev. Lett.},
  volume = {134},
  issue = {1},
  pages = {016401},
  numpages = {7},
  year = {2025},
  month = {Jan},
  publisher = {American Physical Society},
  doi = {10.1103/PhysRevLett.134.016401},
  url = {https://link.aps.org/doi/10.1103/PhysRevLett.134.016401}
}

@article{onsager_reciprocal_1931,
	title = {Reciprocal {Relations} in {Irreversible} {Processes}. {I}.},
	volume = {37},
	copyright = {http://link.aps.org/licenses/aps-default-license},
	issn = {0031-899X},
	url = {https://link.aps.org/doi/10.1103/PhysRev.37.405},
	doi = {10.1103/PhysRev.37.405},
	number = {4},
	urldate = {2025-05-05},
	journal = {Physical Review},
	author = {Onsager, Lars},
	month = feb,
	year = {1931},
	pages = {405--426},
}

@article{Togo2024,
author = {Atsushi Togo and Kohei Shinohara and Isao Tanaka},
title = {Spglib: a software library for crystal symmetry search},
journal = {Science and Technology of Advanced Materials: Methods},
volume = {4},
number = {1},
pages = {2384822},
year = {2024},
publisher = {Taylor \& Francis},
doi = {10.1080/27660400.2024.2384822},
URL = { https://doi.org/10.1080/27660400.2024.2384822},
eprint = { https://doi.org/10.1080/27660400.2024.2384822}
}

@article{Iraola2022,
  author  = {Iraola, Mikel and Ma{\~n}es, Juan L. and Bradlyn, Barry and Neupert, Titus and Vergniory, Maia G. and Tsirkin, Stepan S.},
  title   = {{IrRep}: Symmetry eigenvalues and irreducible representations of ab initio band structures},
  journal = {Computer Physics Communications},
  volume  = {272},
  pages   = {108226},
  year    = {2022},
  doi     = {10.1016/j.cpc.2021.108226}
}

@article{Ong2013,
  title = {Python Materials Genomics (pymatgen): A robust, open-source python library for materials analysis},
  author = {Ong, Shyue Ping and Richards, William Davidson and Jain, Anubhav and Hautier, Geoffroy and Kocher, Michael and Cholia, Shreyas and Gunter, Dan and Chevrier, Vincent L. and Persson, Kristin A. and Ceder, Gerbrand},
  journal = {Computational Materials Science},
  volume = {68},
  pages = {314--319},
  year = {2013},
  doi = {10.1016/j.commatsci.2012.10.028},
  url = {https://www.sciencedirect.com/science/article/abs/pii/S0927025612006295}
}

@misc{prodencerGitHub,
  author={Buiarelli, Luca and Jung, Seongjoo and Park, Hyeonseo and Birol, Turan},
  title = {{``ProDenCeR''}},
  year = {2026},
  publisher = {GitHub},
  journal = {GitHub repository},
  url = {https://github.com/lucabuia/prodencer},
}

@article{grimmer_thermoelectric_2017,
	title = {Thermoelectric transport properties in magnetically ordered crystals},
	volume = {73},
	issn = {2053-2733},
	url = {https://journals.iucr.org/paper?S2053273317005368},
	doi = {10.1107/S2053273317005368},
	number = {4},
	urldate = {2025-05-05},
	journal = {Acta Crystallographica Section A Foundations and Advances},
	author = {Grimmer, Hans},
	month = jul,
	year = {2017},
	pages = {333--345},
}

@article{fernandes_loop-current_2026,
	title = {Loop-current order in kagome metals},
	volume = {22},
	issn = {1745-2473, 1745-2481},
	url = {https://www.nature.com/articles/s41567-026-03229-z},
	doi = {10.1038/s41567-026-03229-z},
	number = {7},
	urldate = {2026-08-27},
	journal = {Nature Physics},
	author = {Fernandes, Rafael M. and Birol, Turan and Ye, Mengxing and Vanderbilt, David},
	month = jul,
	year = {2026},
	pages = {991--998},
}

\end{document}